\documentclass[10pt,conference,a4paper]{IEEEtran}
\usepackage{times,amsmath,epsfig}
\usepackage{epstopdf}
\usepackage{multirow}
\usepackage{bm,url}
\usepackage{hyperref}
\title{Neural Network-Based Arithmetic Coding of Intra Prediction Modes in HEVC}
\author{%
Rui Song, Dong Liu, Houqiang Li, Feng Wu\\
\fontsize{10}{10}\selectfont\itshape
CAS Key Laboratory of Technology in Geo-Spatial Information Processing and Application System,\\
University of Science and Technology of China, Hefei 230027, China\\
\fontsize{9}{9}\selectfont\ttfamily\upshape
sruestc@mail.ustc.edu.cn, \{dongeliu,lihq,fengwu\}@ustc.edu.cn
}
\begin{document}
\maketitle

\begin{figure}[b]
\fontsize{8}{9}\selectfont
This work was supported in part by the National Program on Key Basic Research Projects (973 Program) under Grant 2015CB351803, in part by the Natural Science Foundation of China under Grant 61390512, Grant 61325009, and Grant 61425026, and in part by the Fundamental Research Funds for the Central Universities under Grant WK3490000001. \emph{(Corresponding author: Dong Liu.)}
\\[1\baselineskip]
\parbox{\hsize}{\em
IEEE VCIP'17, Dec. 10 - Dec. 13, 2017, St. Petersburg, Florida, USA.

978-1-5386-0462-5/17/\$31.00 \ \copyright 2017 IEEE.
}
\end{figure}

\begin{abstract}
In both H.264 and HEVC, context-adaptive binary arithmetic coding (CABAC) is adopted as the entropy coding method. CABAC relies on manually designed binarization processes as well as handcrafted context models, which may restrict the compression efficiency. In this paper, we propose an arithmetic coding strategy by training neural networks, and make preliminary studies on coding of the intra prediction modes in HEVC. Instead of binarization, we propose to directly estimate the probability distribution of the 35 intra prediction modes with the adoption of a multi-level arithmetic codec. Instead of handcrafted context models, we utilize convolutional neural network (CNN) to perform the probability estimation. Simulation results show that our proposed arithmetic coding leads to as high as 9.9\% bits saving compared with CABAC.
\end{abstract}

\begin{IEEEkeywords}
Arithmetic coding, Context-adaptive binary arithmetic coding, Convolutional neural network, HEVC, Intra prediction mode.
\end{IEEEkeywords}

\section{Introduction}
The High Efficiency Video Coding standard, known as H.265/HEVC \cite{sullivan2012overview}, outperforms its predecessor H.264/AVC \cite{wiegand2003overview} by about 50\% in compression efficiency.
HEVC uses several new tools to improve the coding efficiency, such as prediction and transform with larger block size, new transform tools like discrete sine transform, more intra prediction modes, additional loop filters, and so on.

Context-adaptive binary arithmetic coding (CABAC) \cite{sze2012high} is
an entropy coding method that is applied in both H.264/AVC and H.265/HEVC.
In H.264/AVC, CABAC is supported in the Main and higher profiles of the standard, but context-adaptive variable-length coding (CAVLC) is used in the Baseline profile, since CABAC requires more computations than CAVLC.
Although CABAC incurs higher computational complexity, its compression efficiency is shown to outperform CAVLC by 5\%--15\% \cite{wiegand2003overview}. Thus, CABAC is chosen as the only entropy coding tool in HEVC.

The encoding process of CABAC consists of three steps: binarization, context modeling, and binary arithmetic coding.
If the syntax element is not binary, the encoder will first map the element to a binary sequence.
There are two coding modes: regular and bypass.
For regular coding mode, the probability model of the bin to be encoded is selected by the \emph{context}, which refers to the previously encoded syntax elements.
Then the bin and the selected context model is passed to the arithmetic coding engine, which not only encodes the bin, but also updates the corresponding probability distribution of the context model.
The bypass coding mode is selected for specific bins in order to speed up the entropy coding process with negligible loss of coding efficiency.
In the bypass mode, all bins are encoded with the probability equal to 0.5.

Recently, convolutional neural network (CNN) shows great successes in many computer vision tasks.
More recently, Toderici \emph{et al.} \cite{toderici2017full} proposed an image compression framework using combined CNN and recurrent neural network (RNN).
That framework adopts an RNN to perform the entropy coding of the binary codes.
These works indicate a promising approach to video coding by using neural networks to further improve the compression efficiency.

Motivated by the recent works, we propose an arithmetic coding strategy by training neural networks instead of manually designing binarization and context models. In this paper, we make preliminary studies on coding of the intra prediction modes in HEVC by means of neural network. Our key idea is to train a neural network that predicts the probability distribution of the syntax elements (intra prediction modes in this paper), and then adopt arithmetic codec to encode the syntax elements based on the predicted probability. Even if the syntax element is not binary, we do not perform explicit binarization, but rather adopt multi-level arithmetic coding to cope with.

The remainder of this paper is organized as follows. Section \ref{sec1} presents the
details of the proposed method, including the CNN structure, how to train the CNN, and how to perform arithmetic coding using the CNN.
Section \ref{sec2} gives out the experimental results, followed by conclusions in Section \ref{sec3}.

\section{The Proposed Method}
\label{sec1}
\subsection{CABAC for Intra Prediction Modes in HEVC}
In HEVC intra coding, there are defined two non-directional modes (DC and planar) and 33 directional modes, so there are 35 intra prediction modes in total.
According to the HEVC standard, the binarization for intra prediction modes includes the following steps:
\begin{itemize}
  \item Three most probable modes (MPMs) are derived from the above and the left prediction units (PUs). If the intra prediction mode to be coded is within the MPMs, the MPM flag is set to 1, otherwise set to 0. The MPM flag is first encoded.
  \item Then, if MPM flag is 1, one or two bins are used to indicate the intra prediction mode located in MPMs. If the MPM flag is 0, then there are $35-3=32$ possible modes, so 5 bins are used to encode.
\end{itemize}

After binarization, the MPM flag is encoded using the regular coding engine, which determines the context model according to the MPM flags of the above and the left PUs. The other bins are encoded using the bypass coding mode.

As mentioned before, the CABAC in HEVC has two drawbacks. The first is the explicit binarization step which is manually designed. The second is the context models that are handcrafted. In this paper, we address the two drawbacks by our neural network-based arithmetic coding method.

\subsection{Overview of Our Proposed Method}
In this paper, we propose to use CNN to predict the probability distribution of the syntax elements to be coded, and then use multi-level arithmetic codec to encode the syntax elements based on the predicted probability. It is worth noting that the CNN will take the previously encoded information as input and output the probability prediction, which is equivalent to using context.

\begin{figure}
  \centering
  \includegraphics[width=30mm]{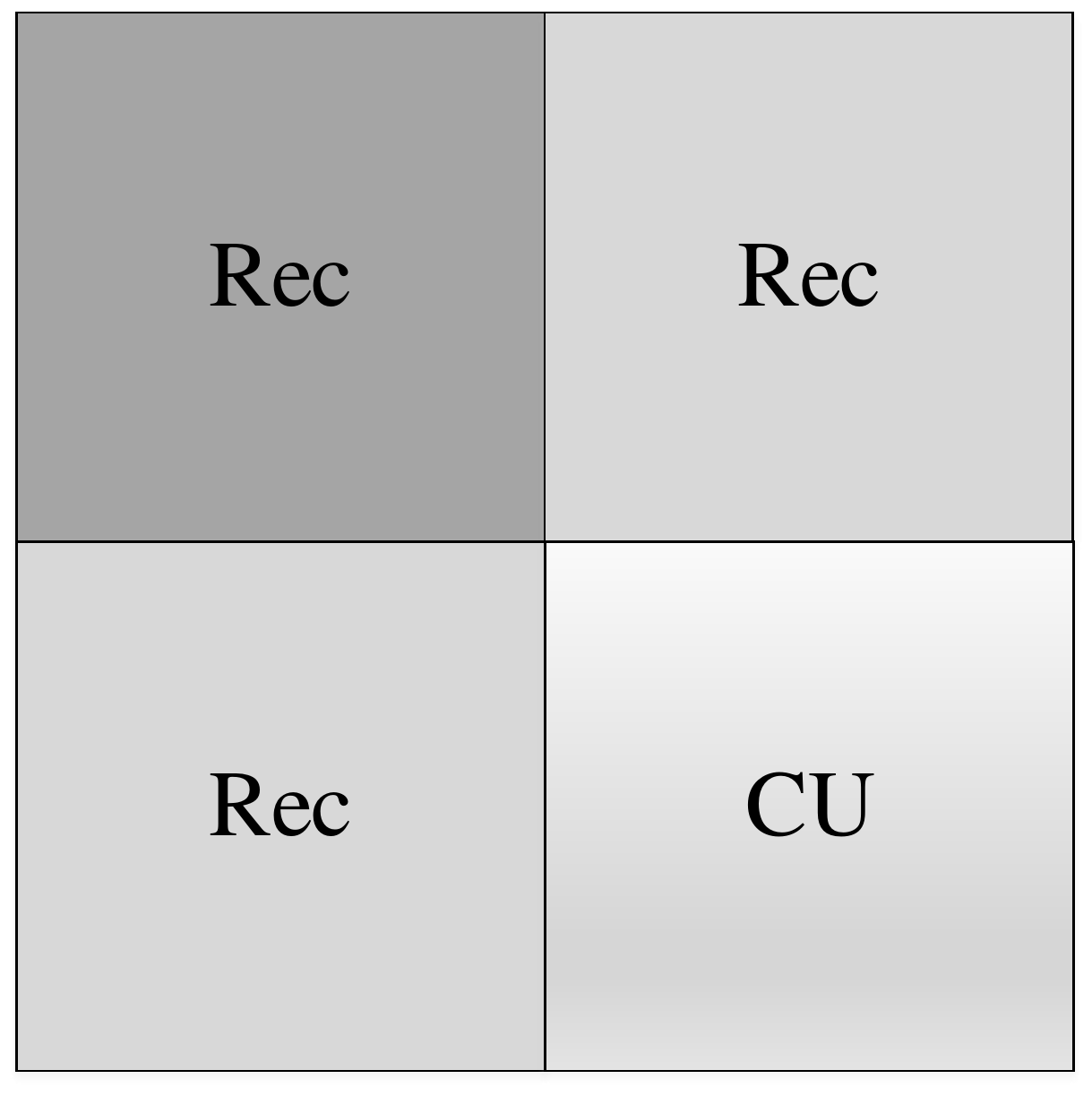}
  \caption{The reconstructed blocks used as inputs to CNN.}\label{rec}
\end{figure}
For intra prediction modes, it is intuitive that the mode to be encoded is highly dependent on the image content. The previously encoded information of image content can be retrieved from the reconstructed blocks. We then propose to use the above-left, the above, and the left reconstructed blocks as inputs to CNN. For simplicity, we use the blocks of the same size as the current block, as shown in Fig. \ref{rec}. In addition, as the MPMs are adopted in HEVC, we also propose to use the derived MPMs as inputs to CNN.

\subsection{Network Structure}
\label{networkstructure}
\begin{figure*}
\centering
  \includegraphics[width=.95\textwidth]{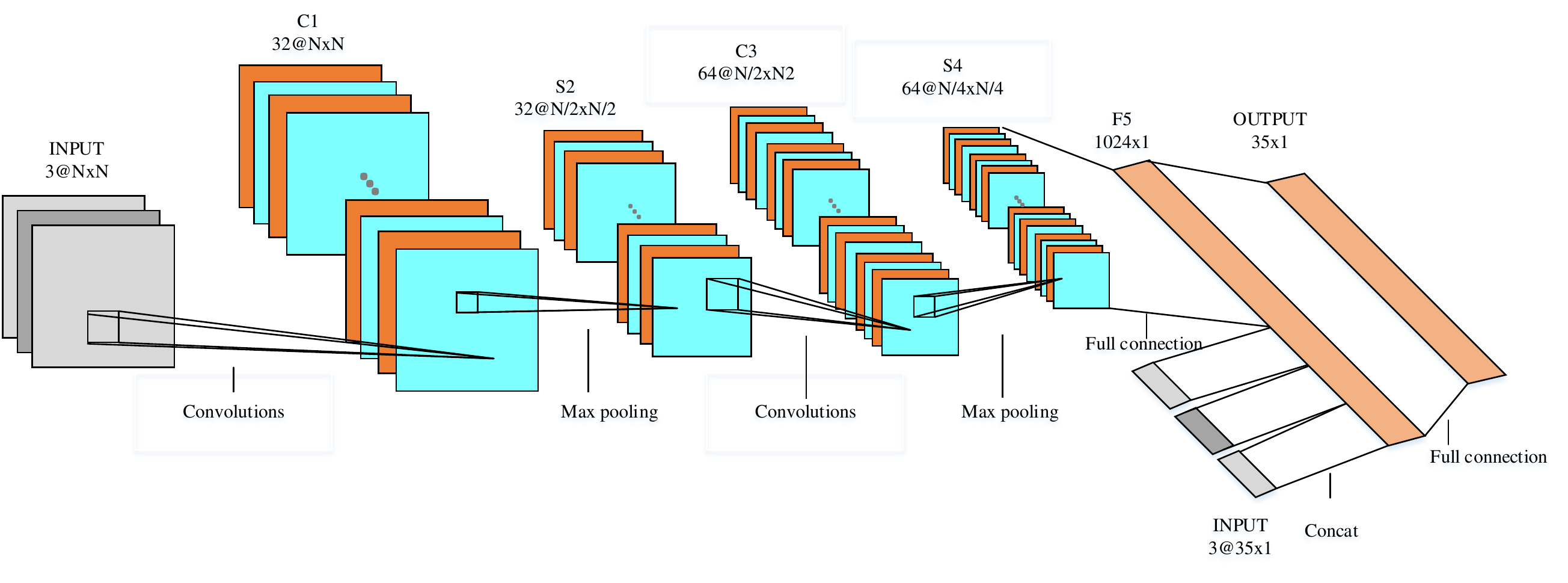}
  \caption{The network structure of our designed CNN.}
  \label{fig:cnn}
\end{figure*}
Our designed network structure is depicted in Fig. \ref{fig:cnn}. Inputs to the network include the neighboring three blocks of size $N\times N$ where $N\times N$ is also the size of the current block to be coded. These blocks are concatenated as three channels, denoted by ``3@$N\times N$'' in the figure. Inputs to the network also include the MPMs derived for the current block. There are three MPMs, each of which is a mode within $\{1,2,\dots,35\}$. For simplicity, each MPM is converted to a 35-dim binary vector, in which there is only one 1 and the other entries are all 0 (known as one-hot vector). There are three 35-dim one-hot vectors, denoted by ``3@35$\times$1'' in the figure. The network outputs a 35-dim vector recording the predicted probability values of the 35 modes.

Our designed network structure is inspired by the well-known LeNet-5 proposed by LeCun \emph{et al.} \cite{lecun1998gradient}, which was originally designed to perform digit recognition from handwritten images. The network structure has been revised according to the settings of our arithmetic coding problem. Specifically, there are two convolutional layers in our network to process the input reconstructed blocks, each of which is followed by a max-pooling layer for down-sampling. The features after the second max-pooling layer are flattened and then mapped into a vector, which is concatenated with the input three one-hot vectors that represent MPMs. One full-connection layer is following to give out the final prediction. The entire network can be summarized by the following equations:
\begin{equation}\label{conv1}
\begin{aligned}
&\mathbf{C}_1 = \mathrm{ReLU}(\mathbf{W}_1*\mathbf{X} + \mathbf{B}_1) \\
&\mathbf{S}_2 = \mathrm{maxpool}(\mathbf{C}_1)\\
&\mathbf{C}_3 = \mathrm{ReLU}(\mathbf{W}_2*\mathbf{S}_2 + \mathbf{B}_2)\\
&\mathbf{S}_4 = \mathrm{maxpool}(\mathbf{C}_3) \\
&\mathbf{F} = \mathrm{ReLU}(\mathbf{W}_3\mathrm{flatten}(\mathbf{S}_4) + \mathbf{B}_3)\\
&\mathbf{F}_5 = \mathrm{concat}(\mathbf{F},\mathbf{MPM}_1,\mathbf{MPM}_2,\mathbf{MPM}_3)\\
&\mathbf{Y} = \mathrm{softmax}(\mathbf{W}_4\mathbf{F}+\mathbf{B}_4)\\
\end{aligned}
\end{equation}
where $\mathbf{X}$ denotes the input reconstructed blocks, $\mathbf{MPM}_i,i=1,2,3$ denote the MPM vectors, and $\mathbf{Y}$ is the output predicted probability. $\mathbf{C}_1,\mathbf{S}_2,\mathbf{C}_3,\mathbf{S}_4,\mathbf{F}_5$ are the internal features, as marked in Fig. \ref{fig:cnn}, $\mathbf{W}_i,\mathbf{B}_i,i=1,2,3,4$ are the weight and bias parameters of the CNN to be learned. $\mathrm{ReLU}$ stands for the rectified linear unit proposed in \cite{nair2010rectified}. $*$ stands for convolution.
$\mathrm{softmax}$ is defined as:
\begin{equation}\label{softmax}
  \mathrm{softmax}(\mathbf{x})|_i=\frac {e^{x_{i}}}{\sum _{k}e^{x_{k}}}
\end{equation}
where $x_i$ or $x|_i$ stands for the $i$-th entry of the vector $\mathbf{x}$.

The detailed configuration of our designed CNN is summarized as follows.
\begin{itemize}
\item For the first convolutional layer, the kernel size is $4\times4$ and there are 32 channels. Zero padding is adopted to ensure the output size is the same to the input size.
\item For the second convolutional layer, the kernel size is also $4\times4$ and there are 64 channels. Zero padding is also adopted.
\item The two max-pooling layers both use kernel size $2\times2$, i.e. down-sampling by a factor of 2.
\item $\mathbf{F}_5$ is 1024-dim, i.e. $\mathbf{F}$ is 919-dim ($1024-35\times3$).
\end{itemize}

\subsection{Training}
To train the CNN that is used for predicting the probability distribution of intra prediction modes, we can use an HEVC compliant encoder to compress some images, from which the probability distribution can be derived. It is worth noting that due to the full-connection layers in our CNN structure (as shown in Fig. \ref{fig:cnn}), the network structure is indeed different for each block size (i.e. $N$). Since HEVC adopts variable block size in prediction and transform, we need to train a different network model for every block size. In this paper we consider two sizes: $N=8$ and $N=16$. For each block size, we want to generate training data as many as possible, so we have revised the HEVC encoder to use fixed block size during encoding. Specifically, we revise the HEVC reference software--HM\footnote{HM version 12.0, \url{https://hevc.hhi.fraunhofer.de/svn/svn_HEVCSoftware/tags/HM-12.0/}.}, setting the maximal coding unit (CU) size to 8$\times$8 (or 16$\times $16), disabling CU quadtree partition and disabling residue quadtree transform. Such encoder is termed HM-intra-8 (HM-intra-16) in the following text.

We use the HM-intra-$N$ ($N$ is 8 or 16) encoder to compress some images. After that, training data are derived from the compressed bitstream. For each $N\times N$ block, indexed by $j$, its actual intra prediction mode is represented as a 35-dim one-hot binary vector, denoted by $\mathbf{T}_j$. Its neighboring reconstructed blocks (as shown in Fig. \ref{rec}) are denoted by $\mathbf{X}_j$, and its derived MPMs are denoted by $\mathbf{MPM}_{ij},i=1,2,3$. Then the network training is driven by minimizing the following loss function:
\begin{equation}\label{loss}
  \mathcal{L}(\Theta) = -\sum_{j}\mathbf{T}_j\cdot \log(\mathbf{Y}_j)
\end{equation}
where $\Theta=\{\mathbf{W}_i,\mathbf{B}_i|i=1,2,3,4\}$ is the parameter set of the network, $\cdot$ stands for inner product.

\subsection{CNN-Based Arithmetic Coding of Intra Prediction Modes}
\label{framework}
\begin{figure}
  \centering
  \includegraphics[width=90mm]{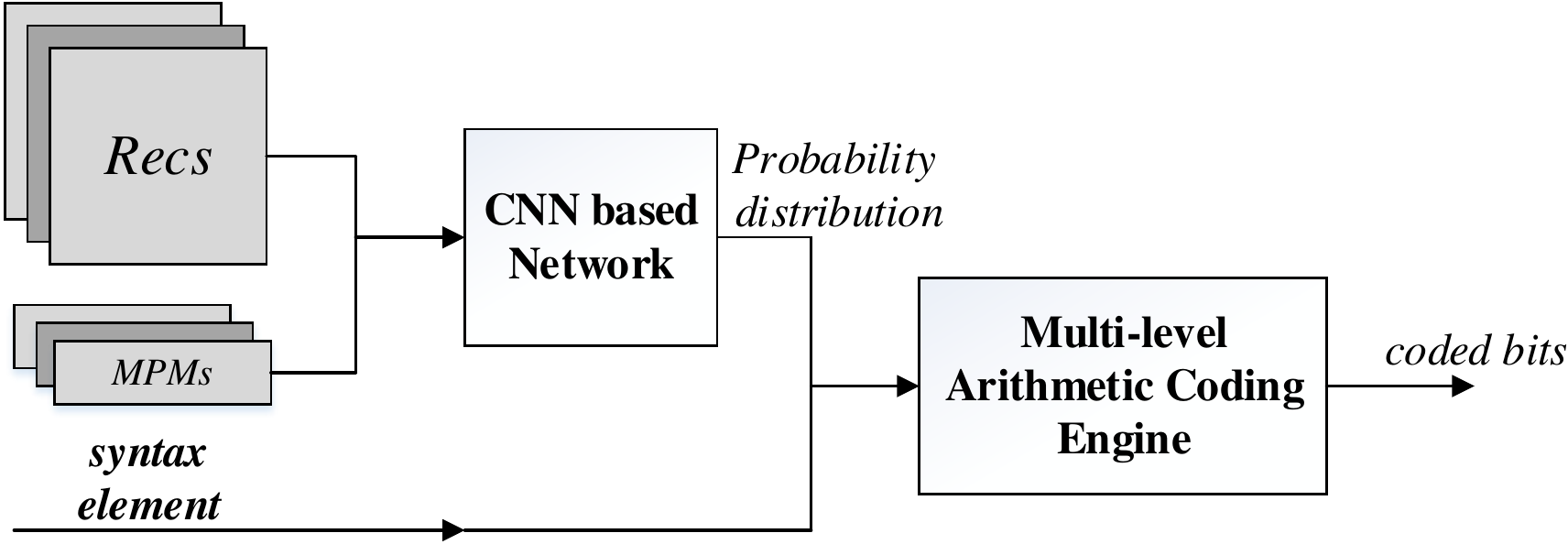}
  \caption{The scheme of CNN-based arithmetic coding.}\label{structure}
\end{figure}

Fig. \ref{structure} presents a generic scheme of CNN-based arithmetic coding. The context information is input into the trained CNN, which predicts the probability distribution for the syntax elements (intra prediction modes in this paper). A multi-level arithmetic coding engine is adopted to encode the syntax elements according to the predicted probability values. We propose to use multi-level arithmetic coding to avoid the binarization step. In this paper, we use the arithmetic codec provided in \cite{said2004introduction}. The decoder is equipped with the identical trained CNN that also predicts the probability, thus the coded bits can be decoded to the corresponding syntax elements.

\section{Experimental Results}
\label{sec2}
\subsection{Training Settings}
We use the deep learning framework TensorFlow\footnote{\url{https://www.tensorflow.org/}} to train
the proposed CNN on an NVIDIA Tesla K40C graphical processing unit.
The loss is minimized using the Adam algorithm \cite{kingma2014adam}.
The training data are obtained from a subset of natural images coming from UCID \cite{schaefer2004ucid},
including 885 images with resolution 512$\times$384. The other images in UCID are used for validation purpose.
We use the abovementioned HM-intra-$N$ encoder to achieve training data for block size $N$. The quantization parameter (QP) is set to 32.
After training, we use the trained CNN together with the HM-intra-$N$ encoder to test the effectiveness of the proposed method.
Test is performed according to HEVC common test conditions with all-intra configuration and QP equal to \{22, 27, 32, 37\}, on the HEVC common test sequences. Please note that we train the CNN with a collection of images from UCID, and test the CNN with the HEVC common test sequences, there is no overlap between training data and test data, so as to demonstrate the generalizability of the network.

\subsection{Comparison with CABAC}
To make a fair comparison with CABAC, we adopt either CABAC or the proposed arithmetic coding method to encode the intra prediction modes, and compare their bits costs. Specifically, we use the abovementioned HM-intra-$N$ encoder to compress each test sequence two rounds. In the first round, the encoder is equipped with CABAC to encode the intra prediction modes, and the total bits cost is denoted by $B_{All1}$. In the second round, the encoder is equipped with the proposed arithmetic coding method, but the other syntax elements such as quantized coefficients are still encoded by CABAC, the bits cost of CABAC is denoted by $B_2$ and the bits cost of our proposed arithmetic coding is denoted by $B_{CNN}$, and the total bits cost is $B_{All2}=B_2+B_{CNN}$. Comparing the two rounds, we can calculate the bits cost of CABAC on intra prediction modes to be $B_{CABAC}=B_{All1}-B_2$. We make comparisons between $B_{CABAC}$ and $B_{CNN}$, and between $B_{All1}$ and $B_{All2}$. It is worth noting that we do not make any change on the mode decision process, i.e. CABAC is always adopted during mode decision. Therefore, both CABAC and our proposed arithmetic coding method are encoding the identical sequences of intra prediction modes. Accordingly, we calculate bits saving by comparing the bits only, since the distortion remains the same.

\begin{table}[t]
	\small
	\caption{Bits savings for intra prediction modes in HM-intra-8}
	\label{table1}
	\centering
	\begin{tabular}{r|rrrr}
		\hline
		QP          &  22            &    27            &    32           &  37                       \\
		\hline
		ClassA             &  --9.9\%       &   --9.8\%        &    --9.6\%      & --8.0\%                            \\
		ClassB             &  --8.9\%       &   --9.1\%        &    --8.7\%      & --6.3\%     \\
		ClassC             &  --10.0\%      &   --10.2\%        &    --9.7\%      & --7.1\%      \\
		ClassD             &  --7.0\%       &   --8.0\%        &    --8.7\%      & --6.6\%      \\
		ClassE             &  --9.7\%       &   --11.5\%        &    --13.0\%      & --12.0\%            \\
		ClassF             &  --8.8\%       &   --9.9\%        &    --9.7\%      & --9.3\%      \\
		Average            &  --9.0\%       &   --9.8\%        &    --9.9\%      & --8.2\%             \\
		\hline
	\end{tabular}
\end{table}

\begin{table}[t]
	\small
	\caption{Overall bits savings in HM-intra-8}
	\label{table2}
	\centering
	\begin{tabular}{r|rrrr}
		\hline
		QP          &  22            &    27            &    32           &  37                       \\
		\hline
		ClassA             &  --0.5\%       &   --0.9\%        &    --1.2\%      & --1.7\%                            \\
		ClassB             &  --0.7\%       &   --1.1\%        &    --1.6\%      & --1.6\%     \\
		ClassC             &  --0.5\%       &   --0.7\%        &    --1.0\%      & --1.0\%      \\
		ClassD             &  --0.3\%       &   --0.5\%        &    --0.9\%      & --1.0\%      \\
		ClassE             &  --0.9\%       &   --1.6\%        &    --2.7\%      & --3.6\%            \\
		ClassF             &  --0.4\%       &   --0.7\%        &    --0.8\%      & --1.0\%      \\
		Average            &  --0.6\%       &   --0.9\%        &    --1.4\%      & --1.6\%             \\
		\hline
	\end{tabular}
\vspace{-12pt}
\end{table}

\begin{table}[t]
	\small
	\caption{Bits savings for intra prediction modes in HM-intra-16}
	\label{table3}
	\centering
	\begin{tabular}{r|rrrr}
		\hline
		QP          &  22            &    27            &    32           &  37                       \\
		\hline
		ClassA             &  --12.5\%       &   --12.2\%        &    --11.3\%      & --9.6\%                            \\
		ClassB             &  --11.3\%       &   --10.1\%        &    --9.0\%      & --6.8\%     \\
		ClassC             &  --11.1\%       &   --11.5\%        &    --9.9\%      & --9.1\%      \\
		ClassD             &  --8.4\%        &   --7.9\%        &    --8.4\%      & --6.5\%      \\
		ClassE             &  --7.1\%        &   --7.2\%        &    --8.9\%      & --7.1\%            \\
		ClassF             &  --5.7\%        &   --6.4\%        &    --7.1\%      & --5.8\%      \\
		Average            &  --9.5\%        &   --9.3\%        &    --9.1\%      & --7.5\%             \\
		\hline
	\end{tabular}
\end{table}

\begin{table}[t]
	\small
	\caption{Overall bits savings in HM-intra-16}
	\label{table4}
	\centering
	\begin{tabular}{r|rrrr}
		\hline
		QP          &  22            &    27            &    32           &  37                       \\
		\hline
		ClassA             &  --0.2\%       &   --0.3\%        &    --0.4\%      & --0.6\%                            \\
		ClassB             &  --0.2\%       &   --0.4\%        &    --0.6\%      & --0.7\%     \\
		ClassC             &  --0.1\%       &   --0.2\%        &    --0.3\%      & --0.5\%      \\
		ClassD             &  --0.1\%       &   --0.1\%        &    --0.2\%      & --0.3\%      \\
		ClassE             &  --0.2\%       &   --0.3\%        &    --0.6\%      & --0.8\%            \\
		ClassF             &  --0.1\%       &   --0.1\%        &    --0.2\%      & --0.2\%      \\
		Average            &  --0.2\%       &   --0.2\%        &    --0.4\%      & --0.5\%             \\
		\hline
	\end{tabular}
\vspace{-12pt}
\end{table}

The comparison results are summarized in Tables \ref{table1} to \ref{table4}.
It can be observed that when CU size is 8$\times$8, using the CNN-based arithmetic coding provides as high as 9.9\% bits saving compared to using CABAC for intra prediction modes, the overall bits savings range from 0.6\% to 1.6\% when QP varies from 22 to 37. At low bit rates (e.g. QP=37), the residue bits are less, and the overall bits saving is relatively higher. When CU size is 16$\times$16, the proposed CNN-based method outperforms CABAC by up to 9.5\%, leading to overall bits savings from 0.2\% to 0.5\%. When CU size is larger, the intra prediction modes are less, and the overall bits saving is relatively lower. It is worth noting that the CNN is trained when QP=32, and results show that the trained CNN can be used and achieve satisfactory results for other QPs.
\section{Conclusion}
\label{sec3}
This paper presents a CNN-based arithmetic coding method for intra prediction modes in HEVC.
We use the CNN to predict the probability distribution of the intra prediction modes, and adopt multi-level arithmetic codec to compress the intra prediction modes with the predicted probability.
Experimental results show that the proposed method can achieve up to 9.9\% bits saving compared with CABAC.
In the future, we will extend the CNN-based arithmetic coding to the other syntax elements in HEVC, such as quantized coefficients and motion vectors.


\end{document}